\documentclass[pra,aps,floats,twocolumn,superscriptaddress,floatfix]{revtex4}
\usepackage{graphicx,amssymb,amsmath,ifthen}
\usepackage[active]{srcltx}
\usepackage[subrefformat=parens,labelformat=parens]{subfig}
\usepackage{lipsum}
\usepackage{float}
\begin{document}
\title{Topological Photonic States at a 1-D Binary-Quaternary Interface}  
\author{Nicholas J. Bianchi}
\affiliation{
  Department of Physics,
  University of Rhode Island,
  Kingston RI 02881, USA}
\author{Leonard M. Kahn}
\affiliation{
  Department of Physics,
  University of Rhode Island,
  Kingston RI 02881, USA}

\begin{abstract}
The existence of topological interface states is investigated at the boundary between a binary photonic crystal and a quaternary photonic crystal, with each  possessing inversion symmetric unit cells. Conditions are established that describe where the quaternary crystal can exist in parameter space subject to constraints. The closing of band gaps is discussed for different optical path ratios. When the binary  and quaternary crystals share an interface, optical  states appear at the interface when the two crystals have different signs for the surface impedance. The evolution of the states is displayed as the geometry of the quaternary crystal changes.
\end{abstract}
\maketitle

%
\section{Introduction}\label{sec:intro}
%
A photonic crystal (PC) is an periodic array of dielectrics and/or conductors able to scatter electromagnetic (EM) fields, where the scattering elements and incident wavelengths are of similar size \cite{Joannopoulos}. This periodicity implies that a PC possesses discrete translational symmetry. Therefore, as in solid state lattices, EM waves are described using  Bloch's Theorem \cite{Bloch}, and the wave equation can be solved for the modes allowed in the PC. Destructive interference due to multiple scattering inside the PC produces frequency ranges in which no mode is allowed to propogate through the crystal regardless of crystal momentum (\textit{i.e.} Bloch wavevector). These regions of surpressed transmission are known as photonic band gaps (PBG), although in 1-D systems they are also called stop bands. While investigations about 1-D photonic systems began with Lord Rayleigh \cite{Rayleigh}, PBG research in two and three dimensions did not accelerate until the works of Yablonovitch  \cite{Yablonovitch1,Yablonovitch2} and John \cite{John} about a century later. Applications of PBGs in PCs are numerous, including dielectric mirrors \cite{Joannopoulos}, channeling EM modes through photonic slab waveguides \cite{Netti} and optical fibers \cite{Knight}, and construction of defect states \cite{Vinck_Posada,Xiao_Qin}. PBGs are also not just limited to PCs; photonic aperiodic structures \cite{Florescu,Jiang,Lei,Poddubny} and certain types of disordered hyperuniform \cite{Torquato} media can also support band gaps, due to isotropy \cite{Man1,Man2,Man3}.

When multiple PCs are joined together, a superlattice can be constructed that can posess optical properties not observed in an isolated crystal. Among these properties is the ability to localize EM fields at an interface. This was theoretically demonstrated by Kavokin \textit{et al.} \cite{Kavokin}, using two adjacent lossless PCs with different periods. The interface states that developed at the boundary of the individual crystals were called optical Tamm states (OTS), due to similarities with electronic surface modes discovered by Tamm \cite{Tamm}, and were found to be strongly dependent on the order of the individual layers in the crystals. Follow up investigations by Vinogradov \textit{et al.} (2006) \cite{Vinogradov1} demonstrated that these OTSs require the normal wavevector to decrease with distance from the interface on both sides. Since there is a PC on both sides, the only way this can happen is if the wave is trying to propagate in a PBG. This implies that a necessary condition for the formation of interface modes is that PBGs of the individual PCs in the superlattice must overlap. These states were experimentally verified soon after and appear as a sharp peak in transmission spectra \cite{Goto}. Kang \textit{et al.} \cite{Kang} used a Bloch Wave Expansion technique for symmetric and asymmetric unit cells to show that OTSs would appear if impedance matching was satisfied at the interface and confirmed the idea that the order of layers in a unit cell mattered for the appearance of states \cite{Kavokin}; however,no physical connection between the appearance of states and the layer order was found \cite{Kang}. While OTSs were mostly studied in asymmetric unit cell configurations,  Vinogradov \textit{et al.} (2010) \cite{Vinogradov2} showed that symmetric unit cells were also valid. Interface states with symmetric unit cells were classified as optical Shockley \cite{Shockley} states (OSSs); however, it was determined that the underlying physical mechanism that produced OTSs and OSSs was the same, thus all optical states are referred to as Tamm states, although many papers simply call them interface states. While the bulk band structure for an infinite crystal will be unaffected by the symmetry of the unit cell, the exact location of the interface state will shift slightly \cite{Vinogradov2}.

Utilizing a symmetric unit cell, a special type of OTS called a topological interface state can be studied. As stated before, the existence of  interface states in a superlattice is strongly dependent on the order of the layers comprising the PCs. Xiao \textit{et al.} \cite{Xiao} was able to explain the surface impedance of a PC in terms of a topological invariant known as the Zak phase \cite{Zak}. A Zak phase is assigned to each isolated bulk band in the band structure. It was shown that interface states emerge in the band gaps if these phases change value via a topological phase transition. One way these phase transitions can occur is if the order of layers in unit cells on one side of the interface is reversed while on the other side it is not. The electric field will acquire a Zak phase (for each band) as the Bloch wavevector, $\kappa$, travels on a closed path around the $1^{\text{st}}$  Brillouin zone. Since the system is 1D, this path is a ring. If the PC is \textit{D}-dimensional, the trajectory that $\kappa$ traces in momentum space would exist on the surface of a \textit{D}-torus.  For a PC whose unit cell possesses inversion symmetry, the Zak phase is a convient measure of topological phase of the band structure as it is constrained to 0 or $\pi$, depending on the inversion center \cite{Zak}. For EM systems with period $\Lambda$, the Zak phase can be written as \cite{Xiao}:

\begin{equation}
\theta^{\text{zak}}_{n}=\! \int_{-\pi/\Lambda}^{\pi/\Lambda} \textit{i} \, \big \langle u_{n,\kappa}\big |\epsilon \, \partial_{\kappa} \big | u_{n,\kappa} \big \rangle  \, d\kappa  \label{eq:zakphase}
\end{equation}
where
\begin{equation}
\textit{i} \, \big \langle u_{n,\kappa}\big |\epsilon \, \partial_{\kappa} \big | u_{n,\kappa} \big \rangle = \textit{i} \! \int_{\text{unit cell}}  u_{n,\kappa}^{*}(z) \, \epsilon(z) \,  \partial_{\kappa}u_{n,\kappa}(z)  \, dz  \label{eq:Berry}
\end{equation}
represents the Berry connection. In  Eq.~\eqref{eq:Berry}, $\epsilon(z)$ is the relative permittivity across the unit cell, and $u_{n,\kappa}(z)$ is the periodic function from Bloch's Theorem, $\textit{E}_{n,\kappa}(z)=\text{exp}(\textit{i} \,\kappa \, z)u_{n,\kappa}(z)$, where  $\textit{E}_{n,\kappa}(z)$ is the electric field. The label $n$ specifies the isolated band.

A topological interface state will appear if the surface impedances of the PCs on both sides of the interface sum to zero \cite{Xiao}:
\begin{equation}
Z_{\text{left}} + Z_{\text{right}}=0  \label{eq:Zstate}
\end{equation}
Since the impedance in a PBG is imaginary, it can be written as $Z/Z_0 = i \zeta$, where:
\begin{equation}
\text{sign}(\zeta^{(n)})=(-1)^{n+l}\text{exp}\left(\textit{i}\sum_{m=0}^{n-1}\theta^{\text{zak}}_{n}\right)  \label{eq:Zsign}
\vspace{5mm}
\end{equation}
with $Z_0$ being the vacuum impedance. The variable \textit{l} indicates the number of band crossings (Dirac points) in the band structure below gap \textit{n}. In a binary PC, a Dirac point will occur if the ratio of the optical path lengths of the two layers is a rational number \cite{Xiao,Nusinsky}. 

After Xiao's  paper, research in this field began to rapidly expand. Experimental measurements of Zak phase in a 1-D $\text{SiO}_2  - \text{TiO}_2$ composite structure were conducted by first measuring reflection phase \cite{Gao1,Gao2}. Other experiments have measured Zak phases through direct obervation of interface states \cite{Wang1}. Recent theoretical work has shown that by manipulating the unit cell inversion centers, a superlattice can be desgined that supports topological states in every PBG\cite{Choi}. The robustness of topological states in photonic systems with a finite number of layers has also been examined as the unit cell number varies \cite{Kalozoumis}. Some photonic systems simultaneously support topological and Fano resonances  \cite{Gao3}. Other works have extended these ideas to include PCs including metallic layers \cite{Deng,Ge,Wang2}.

\begin{figure}[!ht] 
	\centering
	\subfloat[][]{\includegraphics[width=0.8\columnwidth]{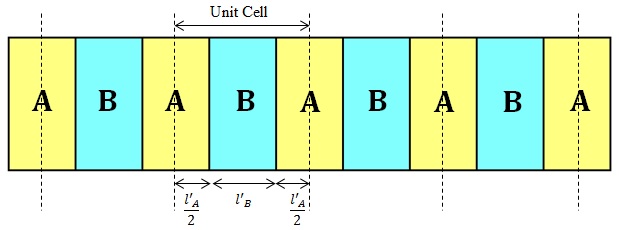}\label{0a}}\\
	\subfloat[][]{\includegraphics[width=0.8\columnwidth]{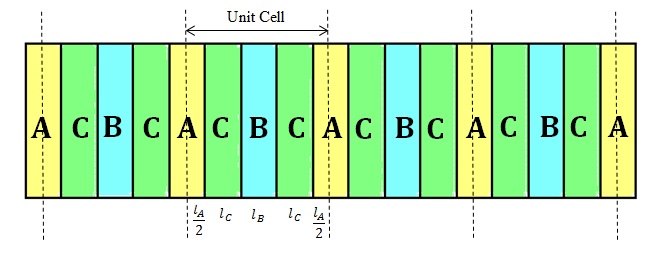}\label{0b}}\\
	\caption{(a) Symmetric unit cells of a binary PC.  (b) Symmetric unit cells of a quaternary PC.}
	\label{fig:PC_examples}
\end{figure}

While the concept of interface states at the boundary of two inversion symmetric binary PCs is well understood, the literature is sparse about what happens when an additional layer is added to one of the crystals \cite{Midya,Li1,Li2}. More specifically, we consider inserting an additional layer inbetween every original layer of the binary PC. The binary crystal is displayed in Fig.~\subref*{0a} while the new crystal is shown in Fig.~\subref*{0b}. Despite being composed of three different materials, this new PC is not ternary. In order to keep the unit cell inversion symmetric, two layer $C$'s must be included, thus giving it four layers and being referred to as a quaternary PC. The primes are used to indicate that the thicknesses of layers $A$ and $B$ can be different between the two PCs, even if the respective layers are composed of the same material. 

In this work, the quaternary PC is created by allowing layer $C$ to expand at every $AB$ and $BA$ interface. When layer $C$ is absent, the superlattice resembles an infinite 1-D binary PC, such as in Fig.~\subref*{1a} (Both crystals are assumed to be semi-infinite. To save space, only the unit cells adjacent to the interface are displayed). As $C$ expands in  Fig.~\subref*{1e}, the crystal on the right hand side (RHS) becomes quaternary. The PC on the left hand side (LHS) always remains binary. Layer $C$ is allowed to expand until layer $B$, $A$ , or both vanish. As will be explained in the next section, at least one other layer type must vanish to preserve the unit cell period. After layer $C$ expands to its maximum side, the quaternary PC becomes binary again but with a different sequence, shown in Figs~\subref*{1c}-\subref*{1d}.

%
\section{Geometry}
%

Before discussing the results of this investigation, it is benifical to establish some dimensionless quantities that will make scaling more natural. Since any periodic PCs are constructed of identical unit cells, we need only consider a single unit cell. The period and optical path length of the unit cell are $\Lambda = l_A + l_B + 2 l_C$ and $\Gamma=n_A l_A + n_B l_B + 2 n_C l_C$, respectively. The individual layers have thicknesses $l_i$ and refractive indices $n_i$, for $i=\{A,B,C\}$. In this paper, $\Lambda$ and  $\Gamma$ are held constant. Therefore, we define a parameter, $\gamma \equiv \Gamma/\Lambda \geq 1$. The layer thicknesses are made dimensionless with $d_i \equiv l_i/\Lambda$.With this information:

\begin{widetext}

\begin{figure}[!ht] 
	\centering
	\subfloat[][]{\includegraphics[width=0.4\columnwidth]{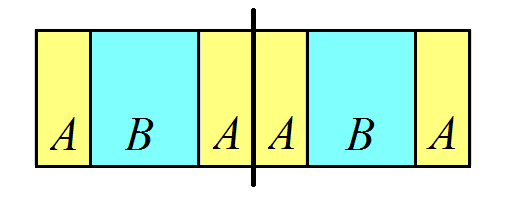}\label{1a}}
	\subfloat[][]{\includegraphics[width=0.4\columnwidth]{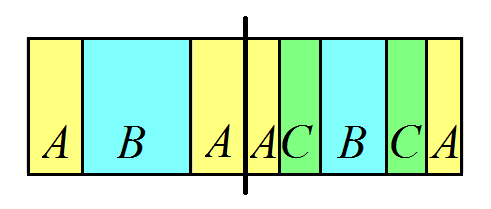}\label{1e}}\\
	\subfloat[][]{\includegraphics[width=0.4\columnwidth]{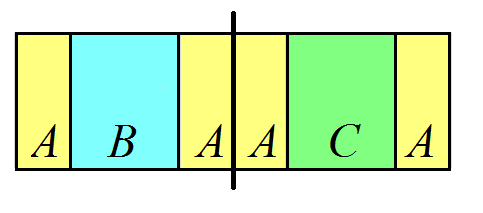}\label{1c}}
	\subfloat[][]{\includegraphics[width=0.4\columnwidth]{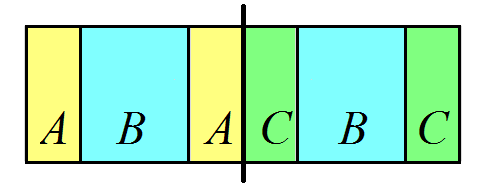}\label{1b}} \\
	\subfloat[][]{\includegraphics[width=0.4\columnwidth]{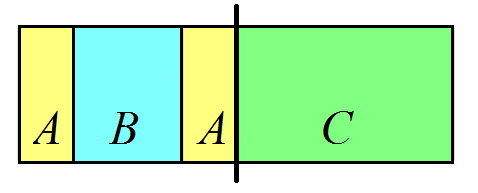}\label{1d}}
	\caption{The interface between two semi-infinite PCs is displayed as a thick black line. On the LHS is a single unit cell from Fig.~\protect\subref*{0a}. On the RHS is a single unit cell from Fig.~\protect\subref*{0b}. (a) $d_C=0$. Both PCs are identical binaries. (b) As $d_C$ increases (while $d_A$ and $d_B$ change to keep $\gamma$ constant), the PC on the RHS becomes quaternary. (c) Final binary configuration on the RHS when  $d_B \rightarrow 0$. (d)  Final binary configuration on the RHS when  $d_A \rightarrow 0$. (e)  Final configuration on the RHS when both  $d_A \rightarrow 0$ and $d_B \rightarrow 0$. The RHS is a semi-infinite uniform medium.}
	\label{fig:PC_diagram}
\end{figure}

\end{widetext}

\begin{equation}
d_A = \frac{\gamma - n_B - 2(n_C - n_B)d_C}{n_A - n_B}   \label{eq:dA}
\end{equation}

\begin{equation}
d_B = \frac{\gamma - n_A - 2(n_C - n_A)d_C}{n_B - n_A}    \label{eq:dB}
\end{equation}

\begin{equation}
d_C =  d_C   \label{eq:dC}
\end{equation}

\begin{widetext}

\begin{figure}[!ht] 
	\centering
	\subfloat[][]{\includegraphics[width=0.3\columnwidth]{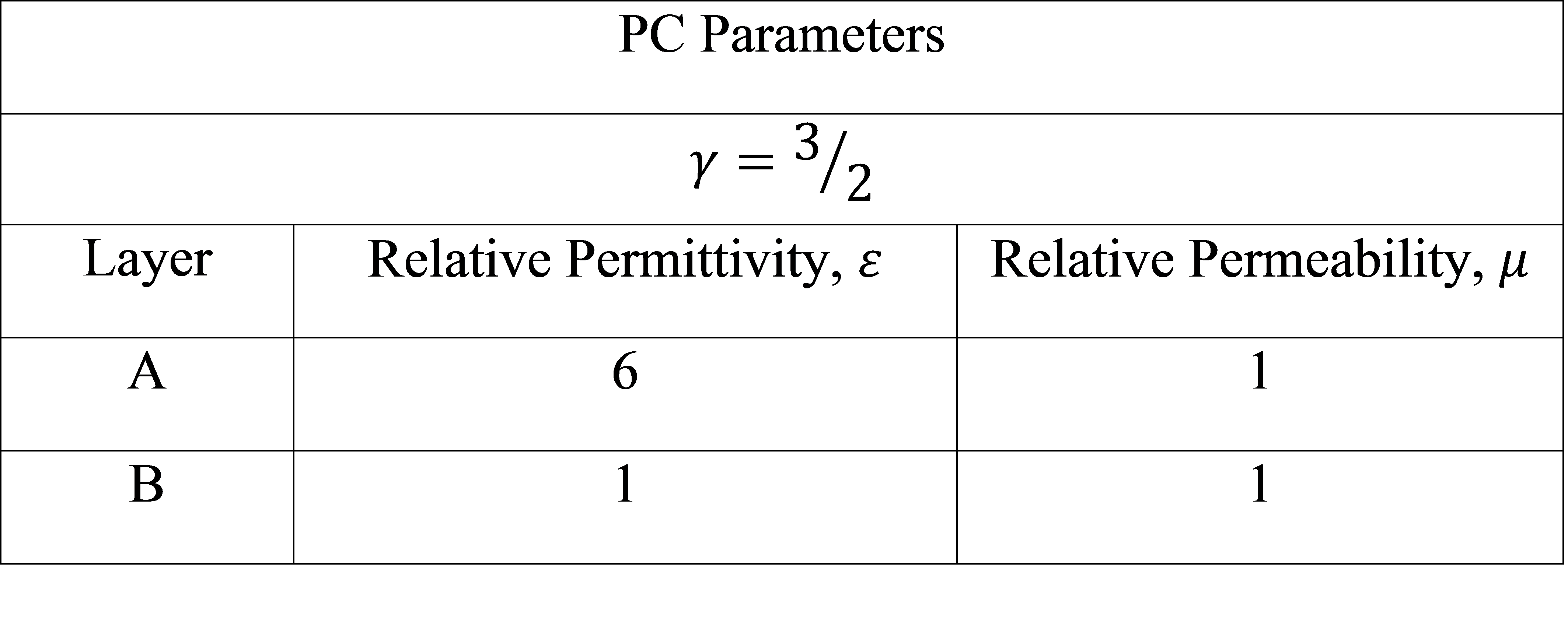}\label{2a}}
	\subfloat[][]{\includegraphics[width=0.3\columnwidth]{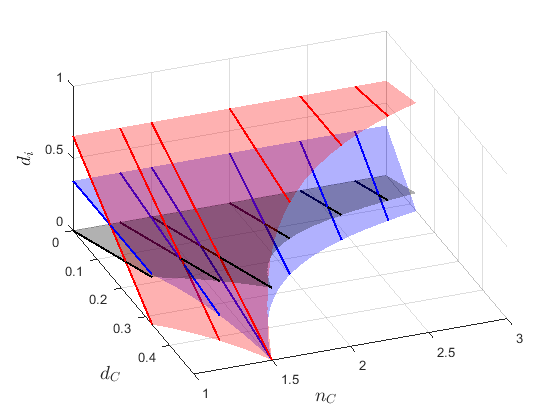}\label{2b}}
	\subfloat[][]{\includegraphics[width=0.3\columnwidth]{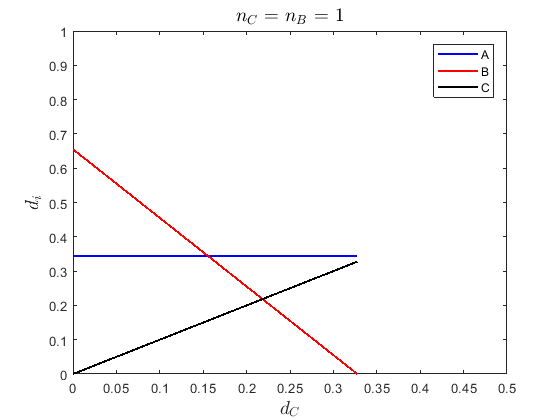}\label{2c}}\\
	\subfloat[][]{\includegraphics[width=0.3\columnwidth]{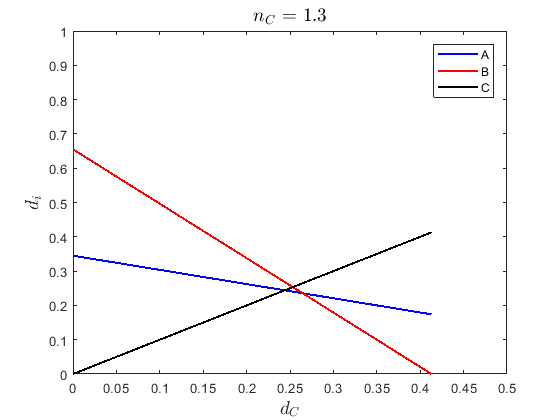}\label{2d}}
	\subfloat[][]{\includegraphics[width=0.3\columnwidth]{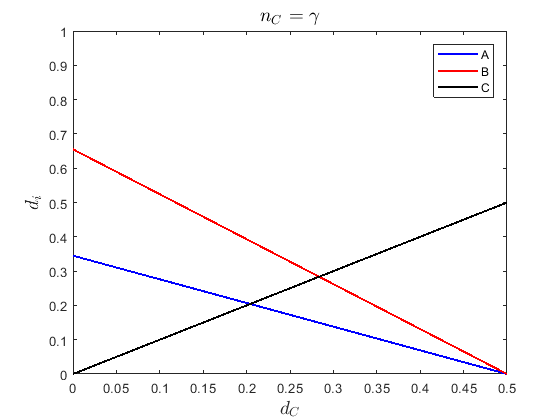}\label{2e}}
	\subfloat[][]{\includegraphics[width=0.3\columnwidth]{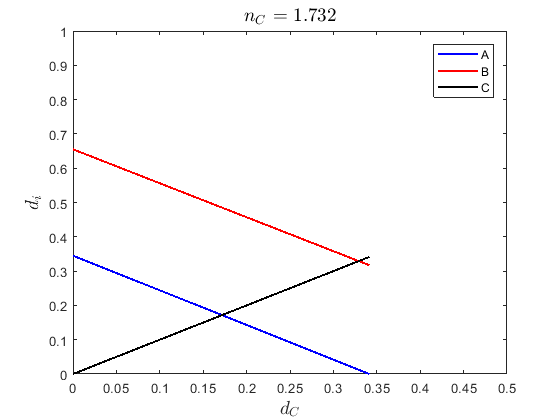}\label{2f}}\\
	\subfloat[][]{\includegraphics[width=0.3\columnwidth]{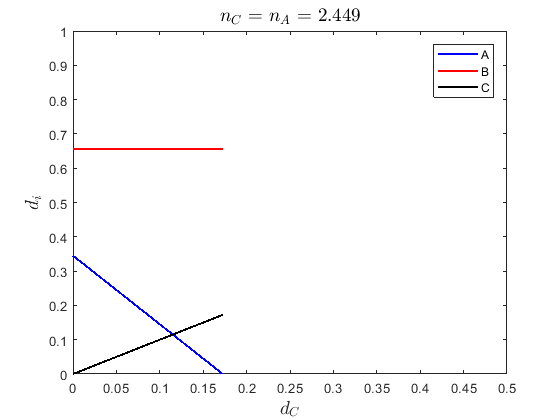}\label{2g}}
	\subfloat[][]{\includegraphics[width=0.3\columnwidth]{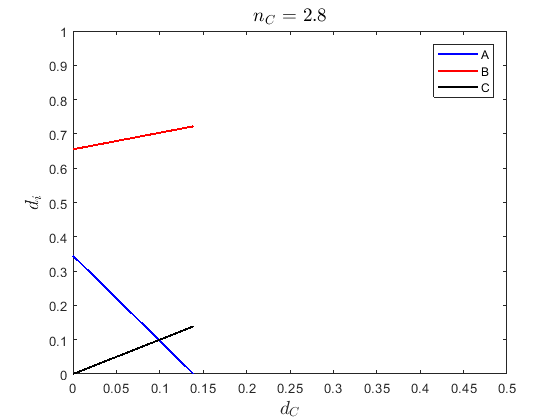}\label{2h}}
	\caption{ (a) $\gamma$, $n_A$, and  $n_B$ are fixed in the system. (b) Surfaces are used to show how the relative thicknesses of layers A (blue), B (red), and C (black) change according to Eq.~\eqref{eq:dA}-\eqref{eq:dC}. Six different trios of lines of constant $n_C$ are highlighted. (c)-(h) Those six trios are plotted seperately as profiles with the value of $n_C$ displayed above each plot. }
	\label{fig:PC_width}
\end{figure}

\end{widetext}

In Eqs.~\eqref{eq:dA} and ~\eqref{eq:dB}, $d_C$ is an independent variable. As $d_C$ changes, $d_A$ and $d_B$ increase or decrease depending on the relative sizes of $\gamma$, $n_A$, and $n_B$. An example of this behavior is shown in Fig. ~\protect\ref{fig:PC_width}. For the given parameters in Fig.~\subref*{2a}, $d_A$, $d_B$, and $d_C$ can exist only on the blue, red, and black surfaces, respectively, shown in Fig.~\subref*{2b}. The bright lines in the surfaces represent behavior of $d_i$ vs $d_C$ for six different values of $n_C$. That is, each $n_C$ value presents a particular vertical slice of Fig.~\subref*{2b}, which is shown in Fig.~\subref*{2c} - ~\subref*{2h}. While each of the six profiles show that, at $d_C=0$, the PC has the binary structure $ABABAB$ (RHS of Fig.~\subref*{1a}), by altering $n_C$, $d_A$ and $d_B$ can display different rates of change with respect to $d_C$. In Figs.~\subref*{2c} and ~\subref*{2d}, $n_C<\gamma$, so as $d_C$ increases, $d_B$ reaches 0 before $d_A$. When this happens, the PC configuration becomes $ACACAC$ (RHS of Fig.~\subref*{1c}). Note that in Fig.~\subref*{2c}, since $n_B = n_C$, $d_A$ is flat \footnote{While $n_B=1$ in this example for simplicity, if $1 < n_B<\gamma$, then $d_A$ would have a positive slope for $n_C < n_B$. The slope would go to 0 as $n_C \rightarrow n_B$}. Also, as $n_C \rightarrow \gamma$, Eqs.~\eqref{eq:dA}-\eqref{eq:dC} can stay positive for larger values of $d_C$. This continues until $n_C = \gamma$, in Fig.~\subref*{2e}. At this point, a cusp is reached and  $d_A = d_B = 0$ when $d_C = 0.5$. Therefore, the PC become a uniform dielectric, $CCCCCC$ (RHS of Fig.~\subref*{1d}). As $n_C$ increases further, in Fig.~\subref*{2f} to $\sqrt{3}$, $d_A$ drops to 0 first. When this occurs, the PC takes the form $CBCBCB$ (RHS of Fig.~\subref*{1b}). The maximum $d_C$ value also starts to decrease. In Fig.~\subref*{2g}, $n_C$ increases until $n_C = n_A$, at which point $d_B$ become flat. Finally, $d_B$ has a positive slope in Fig.~\subref*{2h}, as $n_C > n_A$. 

It is important to reiterate that in the example described in Fig. ~\protect\ref{fig:PC_width}, $n_A$ and $n_B$ are held constant while $n_C$ is free to vary. The surface boundaries in  Fig.~\subref*{2b} are shaped by the conditions that the layer thicknesses must be non-negative and their collective sum must be the period (i.e $d_A+d_B+2d_C=1$). In other words, we can imagine viewing  Fig.~\subref*{2b} from a "top-down" perspective, projecting the surfaces into a 2-D parmeter space of coordinates ($d_C$,$n_C$) In this space, setting Eqs.~\eqref{eq:dA} and ~\eqref{eq:dB} to 0 and solving each for $n_C$ yields:

\begin{equation}
n_C|_{d_A=0} = \frac{\gamma - n_B}{2 d_C} + n_B   \label{eq:n_C_dA}
\end{equation}

\begin{equation}
n_C|_{d_B=0} = \frac{\gamma - n_A}{2 d_C} + n_A   \label{eq:n_C_dB}
\end{equation}

\begin{widetext}

\begin{figure}[!ht]  
	\centering
	\subfloat[][]{\includegraphics[width=0.3\columnwidth]{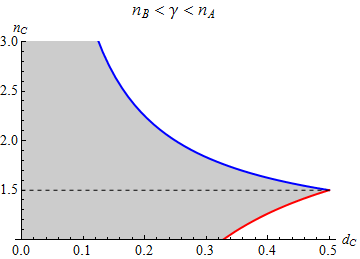}\label{3a}}
	\subfloat[][]{\includegraphics[width=0.3\columnwidth]{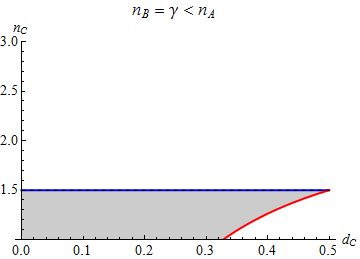}\label{3b}}
	\subfloat[][]{\includegraphics[width=0.3\columnwidth]{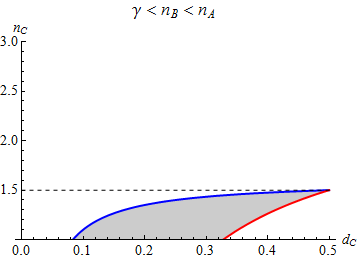}\label{3c}}\\
	\subfloat[][]{\includegraphics[width=0.3\columnwidth]{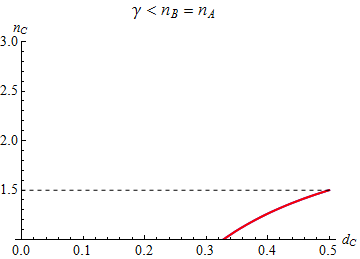}\label{3d}}
	\subfloat[][]{\includegraphics[width=0.3\columnwidth]{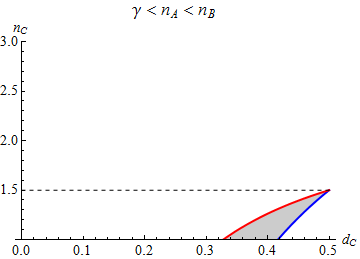}\label{3e}}
	\caption{(a)-(e) The ($d_C$,$n_C$) parameter space is displayed. The PC can only exist in the shaded areas where $d_A$,$d_B$,$d_C$ are all non-negative. $\gamma$ and $n_A$ are constants, with  $\gamma<n_A$  and $n_B$ is free to vary. The dashed line is $n_C=\gamma$. The blue and red lines are Eq.~\eqref{eq:n_C_dA} and Eq.~\eqref{eq:n_C_dB}, respectively. In (d), the shaded region vanishes, since the curves overlap. }
	\label{fig:Parameter_Space_1}
\end{figure}

\begin{figure}[!ht]  
	\centering
	\subfloat[][]{\includegraphics[width=0.3\columnwidth]{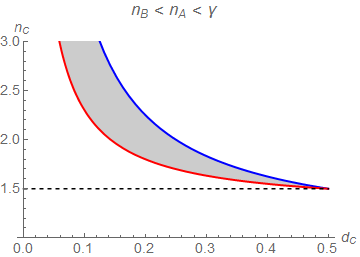}\label{4a}}
	\subfloat[][]{\includegraphics[width=0.3\columnwidth]{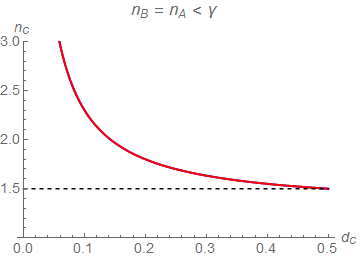}\label{4b}}
	\subfloat[][]{\includegraphics[width=0.3\columnwidth]{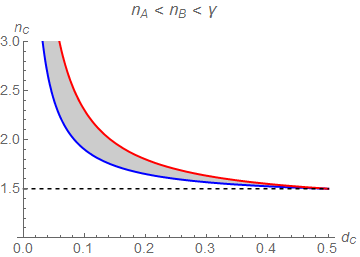}\label{4c}}\\
	\subfloat[][]{\includegraphics[width=0.3\columnwidth]{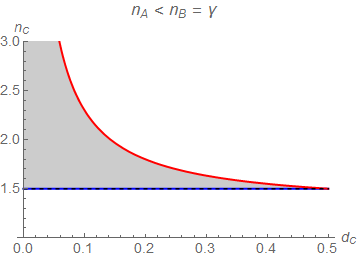}\label{4d}}
	\subfloat[][]{\includegraphics[width=0.3\columnwidth]{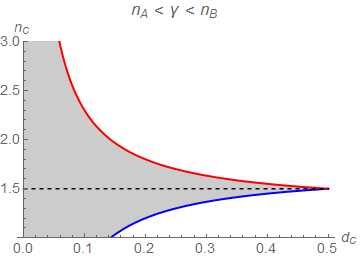}\label{4e}}
	\caption{(a)-(e) Similar to Fig. ~\protect\ref{fig:Parameter_Space_1}, the ($d_C$,$n_C$) parameter space is displayed. The PC can only exist in the shaded areas where $d_A$,$d_B$,$d_C$ are all non-negative. $\gamma$ and $n_A$ are constants, with  $\gamma>n_A$  and $n_B$ is free to vary. The dashed line is $n_C=\gamma$. The blue and red lines are Eq.~\eqref{eq:n_C_dA} and Eq.~\eqref{eq:n_C_dB}, respectively. In (b), the shaded region vanishes, since the curves overlap}
	\label{fig:Parameter_Space_2}
\end{figure}

\end{widetext}

Eqs.~\eqref{eq:n_C_dA} and ~\eqref{eq:n_C_dB} are displayed in Fig. ~\protect\ref{fig:Parameter_Space_1} and represent the boundaries in parameter space. The blue curve is Eq.~\eqref{eq:n_C_dA} and the red curve is Eq.~\eqref{eq:n_C_dB}. The shaded area is the region where the PC is allowed to exist.  Here $\gamma < n_A$ and both constants are given the same values they had in  Fig.~\subref*{2a}; however, now $n_B$ is allowed to change. In  Fig.~\subref*{3a}, since $n_B < \gamma$, the shaded region is unbounded as $n_C \rightarrow \infty$, although Eq.~\eqref{eq:n_C_dA} asymptotically approaches $d_C = 0$. Fig.~\subref*{3a} represents the parameter space for Fig.~\subref*{2b}. As $n_B \rightarrow \gamma$, the shaded area transforms into Fig.~\subref*{3b}. Now there is a strict upper bound on $n_C$, above which the PC cannot exist. Further increasing $n_B$ causes ~\eqref{eq:n_C_dA} to become concave-down. In Fig.~\subref*{3c}, this cuts off $d_C = 0$ from the available region. This means that in order for the PC to exist, it must contain layer $C$ everywhere. As $n_B \rightarrow n_A$, in Fig.~\subref*{3d}, the two curves overlap, causing the region to vanish. It then opens again in Fig.~\subref*{3e}, but with Eq.~\eqref{eq:n_C_dA} ahead of Eq.~\eqref{eq:n_C_dB} \footnote{If $\gamma$ and $n_B$ were constants, with  $\gamma<n_B$ and $n_A$ allowed is change, then the positions of the red and blue curves would be reversed in each plot of Fig. ~\protect\ref{fig:Parameter_Space_1}.}.

Fig. ~\protect\ref{fig:Parameter_Space_2} displays a similar situation to Fig. ~\protect\ref{fig:Parameter_Space_1} except now $n_A < \gamma$. The main difference is that $n_C = \gamma$ serves as a lower bound rather than an upper bound.

%
\section{Results}
%

Now that the conditions for the existence of an inversion-symmetric 4-layer PC and its connection to the binary crystal have been shown, it is easier to discuss the conditions in which a topological state can form. By using the transfer matrix method (See Appendix), we can extend the familiar transfer matrix elements from a binary unit cell \cite{Xiao,Yariv}:

\begin{equation}
t_{11} = e^{i \phi_A}\left(\cos\phi_B + i z^{+}_{AB}\sin\phi_B \right)   \label{eq:t11_2}
\end{equation}

\begin{equation}
t_{12} = i e^{-i \phi_A}  z^{-}_{AB} \sin\phi_B  \label{eq:t12_2}
\end{equation}
to a symmetric 4 layer unit cell:
\begin{widetext}
\begin{equation}
\begin{split}
t_{11} &= e^{i \phi_A} \Big[ \cos\phi_B \cos(2 \phi_C) + \left(i z^{+}_{AC} \cos\phi_B - z^{+}_{BC} \sin\phi_B \right) \sin(2 \phi_C) \\
           &+ i\left( z^{+}_{AB} \cos^{2}\phi_C - z^{+}_{ABC} \sin^{2}\phi_C   \right) \sin\phi_B  \Big]   \label{eq:t11_4}
\end{split}
\end{equation}

\begin{equation}
\begin{split}
t_{12} &= i e^{-i \phi_A} \Big[ z^{-}_{AC} \cos\phi_B \sin(2 \phi_C) + \left(z^{-}_{AB} \cos^{2}\phi_C - z^{-}_{ABC} \sin^{2}\phi_C \right)\sin\phi_B \Big]   \label{eq:t12_4}
\end{split}
\end{equation}
\end{widetext}
where $\phi_i \equiv k_i l_i = 2\pi n_i d_i \xi$ for dimensionless frequency $\xi=f \Lambda/c_0$. $f$ is the  frequency and $c_0$ is the speed of light in vacuum. The impedance mismatch terms are defined as:
\begin{equation}
 z^{\pm}_{ij}\equiv \frac{1}{2}\left( \frac{z_i}{z_j} \pm \frac{z_j}{z_i} \right)    \label{eq:mismatch1}
\end{equation}
\begin{equation}
z^{\pm}_{ABC}\equiv \frac{1}{2}\left( \frac{z_A z_B}{z^2_C} \pm \frac{z^2_C}{z_A z_B} \right)   \label{eq:mismatch2}
\end{equation}
for relative impedance, $z_i$.
It is easy to show that Eq.~\eqref{eq:t11_4} reduces to Eq.~\eqref{eq:t11_2} and Eq.~\eqref{eq:t12_4} reduces to Eq.~\eqref{eq:t12_2} when either $d_B \rightarrow 0$ or $d_C \rightarrow 0$. A third case, $d_A \rightarrow 0$, will also work, but is a bit more subtle; in order for this case to simplify, the substitution $z_A \rightarrow z_C$ must be made. Adding Eq.~\eqref{eq:t11_4} with its complex conjugate yields the dispersion relation  (See Appendix):
\begin{equation}
\begin{split}
\cos(\kappa \Lambda)& = \cos\phi_A \cos\phi_B \cos(2\phi_C)\\
                                   & - \left(z^{+}_{BC}\cos\phi_A \sin\phi_B + z^{+}_{AC}\sin\phi_A \cos\phi_B \right)\sin(2\phi_C)\\
			   &+ \left(z^{+}_{ABC} \sin^2\phi_C - z^{+}_{AB} \cos^2\phi_C \right) \sin\phi_A \sin\phi_B     \label{eq:band_structure}
\end{split}
\end{equation}

With Eqs.~\eqref{eq:t11_4} and \eqref{eq:t12_4}, we can use the expression for surface impedance in Ref.~ \cite{Xiao}, assuming the interface separating the binary and quaternary PCs is at $z=0$:
\begin{equation}
\text{Im}\frac{Z}{Z_0} = z_A\frac{t_{12}\exp(i\phi_A)+\left(\exp(i \kappa \Lambda)-t_{11}\right)}{t_{12}\exp(i\phi_A)-\left(\exp(i \kappa \Lambda)-t_{11}\right)}  \label{eq:Z_equation}
\end{equation}
Note that Eq.~\eqref{eq:Z_equation} is simply the ratio of Eqs.~\eqref{eq:Efield} and ~\eqref{eq:Hfield} with eigenvector components: $a^{(1)}_0=t_{12}$ and $b^{(1)}_0=\exp(i \kappa \Lambda)-t_{11}$ \cite{Yariv}. Inserting  Eq.~\eqref{eq:Z_equation} into Eq.~\eqref{eq:Zstate}, gives a complete equation for a topological interface state. By letting $\kappa \Lambda = n\pi+ix$ \cite{Yariv}, the Bloch phase in Eq.~\eqref{eq:Z_equation} can be re-written in terms of the dispersion relation:
\begin{equation}
e^{i \kappa \Lambda} = (-1)^n e^{-x}
\end{equation}
where,
\begin{equation}
\cosh(x) = (-1)^{-n} \cos(\kappa \Lambda)
\end{equation}
The value $n$ denotes the band gap number and $x$ represents a decay factor.

\begin{figure}[!ht]  
	\centering
	\subfloat[][]{\includegraphics[width=0.8\columnwidth]{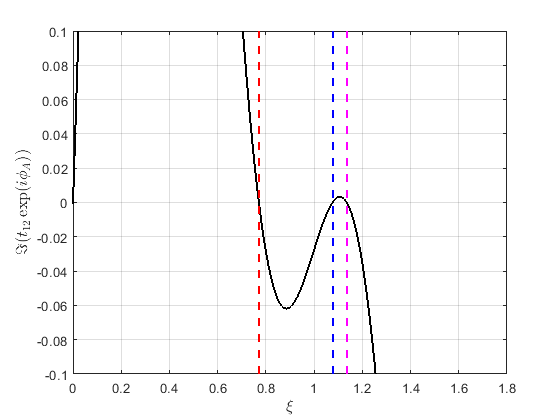}\label{5a}}\\
	\subfloat[][]{\includegraphics[width=0.8\columnwidth]{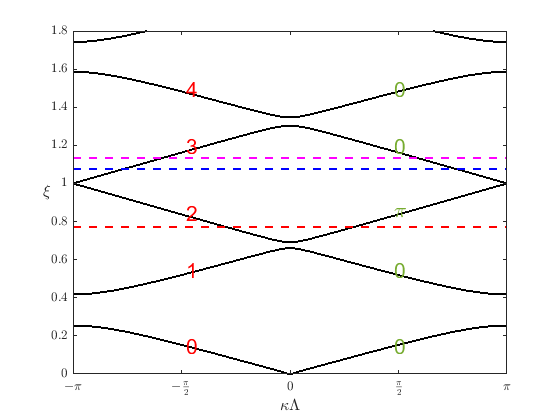}\label{5b}}
	\caption{ System Parameters: $d_A=0.2005$, $\epsilon_A=1$, $\mu_A = 6$, $d_B=0.5135$, $\epsilon_B=1$, $\mu_B = 1$, $d_C=0.143$, $\epsilon_C=1$, $\mu_C = 3$~ (a) $\text{Im}(t_{12}\exp(i\phi_A))$ is plotted with respect to frequency, $\xi$. The zeros are displayed as vertical dashed lines. (b) Band structure is plotted. Those three frequencies are superimposed on the plot. The band numbers are displayed in red and the Zak phases in green.  There is a very thin PBG at $\xi \approx 0.9996$.}
	\label{fig:DoubleCrossing}
\end{figure}

As in Ref.~ \cite{Xiao}, calculating the Zak phase for each isolated band, from Eq.~\eqref{eq:band_structure}, requires finding the set of frequencies, $\xi$, in which $\text{Im}(t_{12}\exp(i\phi_A))=0$, assuming the center of layer $\textit{A}$ is chosen as the center of inversion. If such a value of $\xi$ intersects a band $n>0$, then for that band, $\theta^{\text{zak}}_{n}=\pi$; for all bands not intersected,  $\theta^{\text{zak}}_{n}=0$. For the binary PC, the $\xi$ and thus the Zak phases can be found analytically for all bands. This is done using Eq.~\eqref{eq:t12_2} \cite{Xiao}: $z^{-}_{AB} \sin\phi_B=0$. For the $0^{\text{th}}$ band:
\begin{equation}
\exp(i\theta^{\text{zak}}_0)=\text{sign}(z^{-}_{AB}) \label{eq:band0_2}
\end{equation}
For all other bands, $\sin\phi_B=0$. A similar procedure can be done for the 4-layer unit cell, using Eq.~\eqref{eq:t12_4}; however, one quickly realizes that now $\xi$ cannot be found analytically. Furthermore, the situation is complicated by the fact that $\theta^{\text{zak}}_0$ cannot be separately calculated.  While it is still true that  $\theta^{\text{zak}}_{n>0}=\pi$ for bands intersected by $\xi$, this rule does not appear to consistently hold for the $0^{\text{th}}$ band. In addition, for bands $n>0$, there may be instances where two different $\xi$ values intersect the same band. If this happens, that band has  $\theta^{\text{zak}}_{n>0}=0$; two crossings are treated as no crossing. An example of this behavior is displayed in Fig. ~\protect\ref{fig:DoubleCrossing}. Note that in Fig.~\subref*{5a}, the local maximum would be equal to zero for $d_C \approx 0.1425821$. In that case, the blue and magenta $\xi$ values would become repeated solutions, and despite there only being one root, $\theta^{\text{Zak}}_3$ would still be 0 since an infintesimal increase or decrease in $d_C$ would result in a double crossing or no crossing, respectively. The dashed lines in Fig.~\subref*{5a} are plotted with the band structure in Fig.~\subref*{5b}. As $d_C$ increases, the blue $\xi$ will shift down. When $d_C \approx 0.1461$, bands 2 and 3 will cross, closing the gap. At that moment, the blue line will exist exactly where the bands cross, at $\xi \approx 0.9996$. Increasing $d_C$ further reopens the gap, but now the blue line has moved down to the $2^{\text{nd}}$ band. The $2^{\text{nd}}$ band would now have a double crossing with $\theta^{\text{zak}}_2 =0$ and the $3^{\text{rd}}$ band would now have a single crossing with $\theta^{\text{zak}}_3 =\pi$. 

\begin{widetext}

\begin{figure}[!ht]  
	\centering
	\subfloat[][]{\includegraphics[width=0.4\columnwidth]{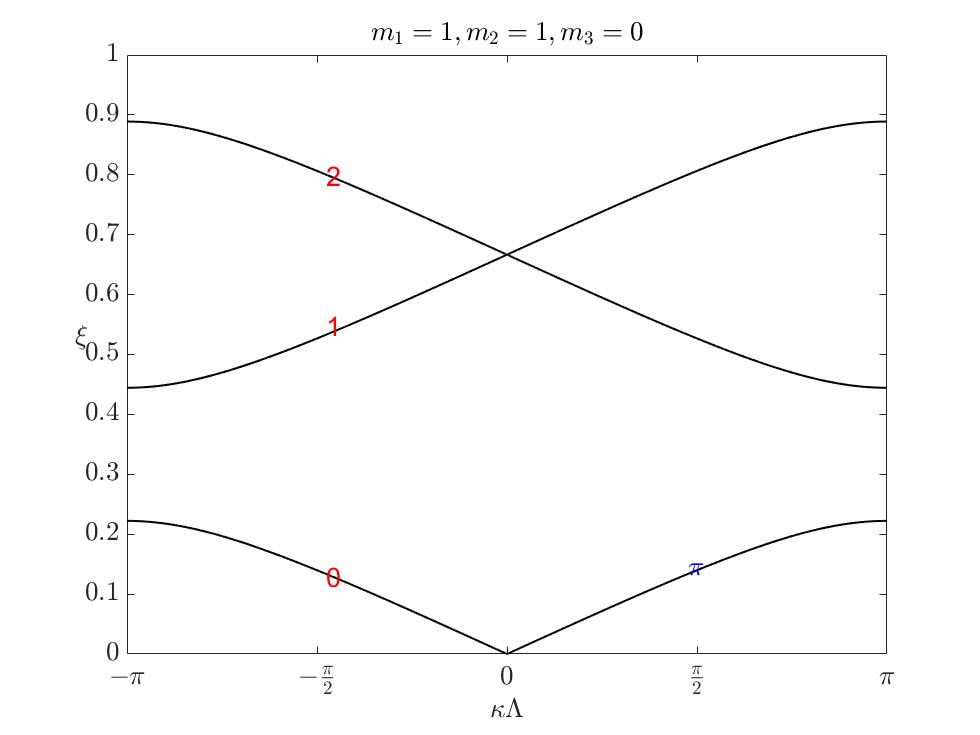}\label{6a}}
	\subfloat[][]{\includegraphics[width=0.4\columnwidth]{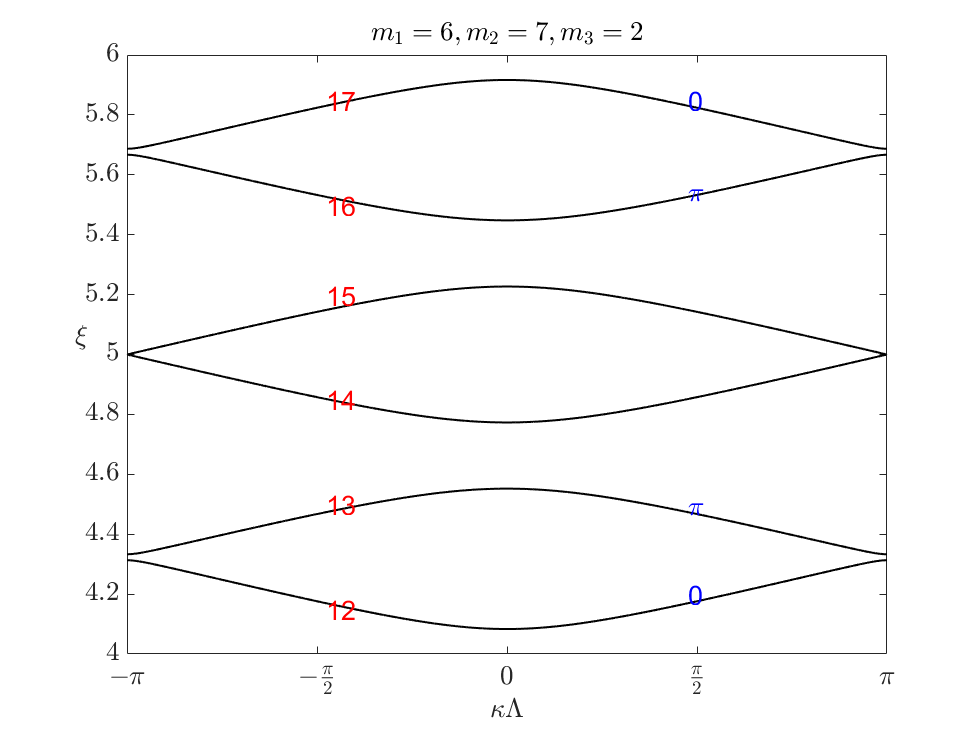}\label{6b}}\\
	\subfloat[][]{\includegraphics[width=0.4\columnwidth]{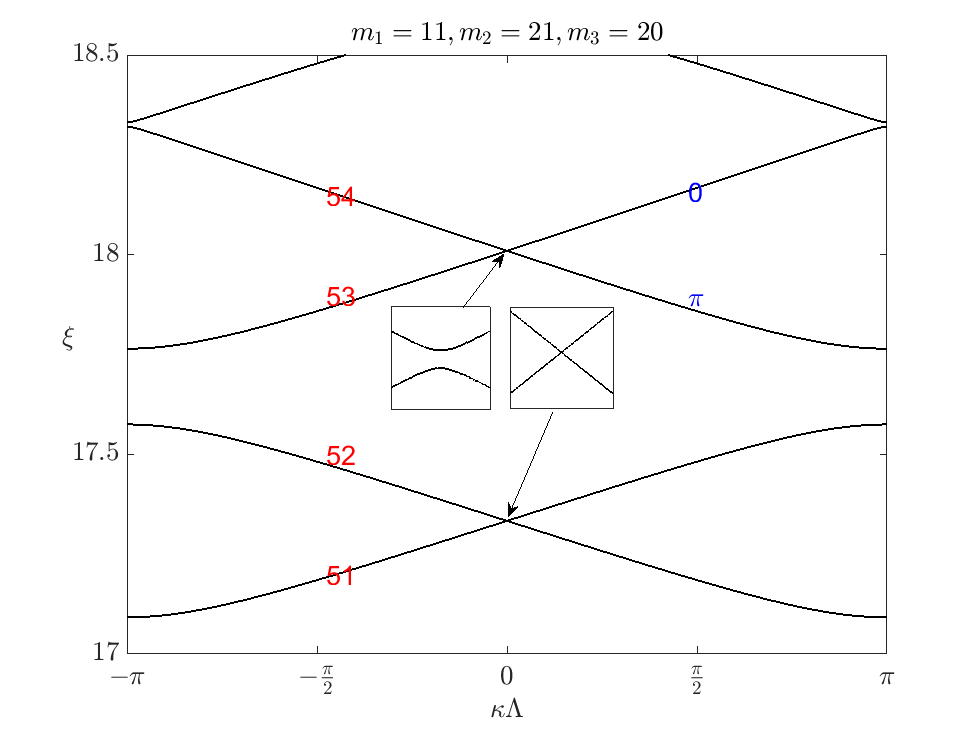}\label{6c}}
	\subfloat[][]{\includegraphics[width=0.4\columnwidth]{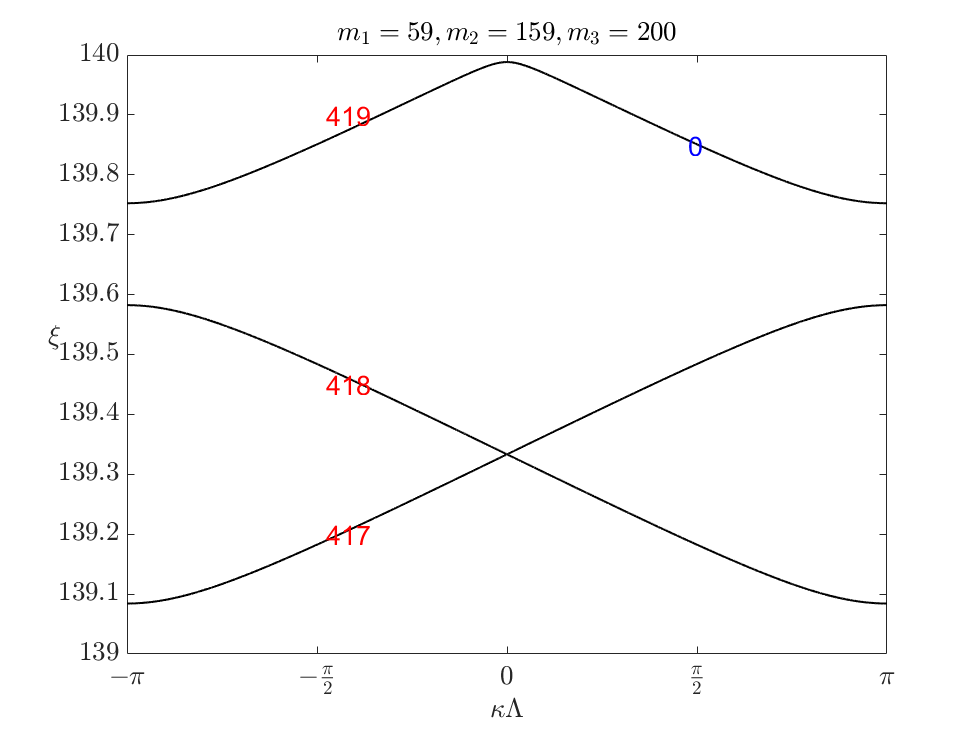}\label{6d}}\\
	\caption{ Band crossing examples for  $\phi_B = M\phi_C$ and $l=1$ (lowest crossing) For each case, $\gamma=1.5$, $n_A=3$, $n_B=1$, $n_C=2$. Red values are band numbers and blue values are Zak phases. Only isolated bands have a Zak phase. (a) $M=\infty$   This is just a binary PC. The first crossing occurs between bands 1 and 2.  (b) $M = 7$ The first crossing occurs between bands 14 and 15. (c) $M = 2.1$  The first crossing occurs between bands 51 and 52. The two inserts show that it can be difficult to distingish a crossing from a thin gap solely by eye.   (d) $M = 1.59$   The first crossing occurs between bands 417 and 418.} 
	\label{fig:Crossing}
\end{figure}

\end{widetext}

Analytic results for band crossings can be obtained if the constraint, $M \phi_C = \phi_B$, is applied, for $M \in \mathbb{Q}$, assuming that all layer widths remain non-negative. Applying this condition to Eqs.~\eqref{eq:dA} - ~\eqref{eq:dC} yields:
\begin{equation}
d_A = \frac{(M+2)n_B n_C - \gamma(2n_B + M n_C)}{(M+2)n_B n_C - n_A(2n_B + M n_C)}   \label{eq:dA_M}
\end{equation}
\begin{equation}
d_B = \frac{M n_C(\gamma - n_A)}{(M+2)n_B n_C - n_A(2n_B + M n_C)}   \label{eq:dB_M}
\end{equation}
\begin{equation}
d_C = \frac{n_B(\gamma - n_A)}{(M+2)n_B n_C - n_A(2n_B + M n_C)}   \label{eq:dC_M}
\end{equation}
It is easy to check that as $M \rightarrow \infty$, $d_C \rightarrow 0$ while $d_A$ and $d_B$ reduce to their respective binary expressions.
This constraint allows for the following band crossing condition to hold \cite{Li1,Li2}\footnote{The factor of 2 in the last term is due to the total length of layer $C$ being $2d_C$. In \cite{Li1,Li2}, there is only a single layer $C$, hence no extra factor}:
\begin{equation}
n_A d_A : n_B d_B : 2 n_C d_C = m_1 : m_2 : m_3      \label{eq:ratio_3}
\end{equation}
for $\{m_1,m_2 ,m_3\} \in\mathbb{N}$. Therefore, bands $l (m_1 + m_2 + m_3)$ and $l (m_1 + m_2 + m_3)-1$ will cross at frequency $\xi_{\text{cross}} = l (m_1 + m_2 + m_3)/(2\gamma)$, where $l\in\mathbb{N^{+}}$. It is productive to illustrate these crossing with examples. In Fig.~\protect\ref{fig:Crossing}, four examples of band crossings are shown, each with a different $M$ value. To ensure that a crossing exists, all refractive indices are rational numbers. Therefore, Eq.~\eqref{eq:ratio_3} can be written as a trio of non-negative integers. In Fig~\subref*{6a}, when one of the terms is 0, the PC becomes binary and the first crossing occurs at a low frequency. In fact, for the particular refractive index values used in this example, only the $0^{\text{th}}$ band is isolated. Fig~\subref*{6b}-~\subref*{6d}, show that as the number of non-repeating digits in $M$ increases, the first crossing occurs at ever higher band numbers, tending to $\infty$ as $M$ becomes irrational. Note that in Fig~\subref*{6b}, the lower band in the crossing pair is even, leading to the crossing happening at the band edge; in the other three plots, the lower band in the crossing pair is odd, so the crossing occurs at the band center. Fig~\subref*{6c} also highlights the difficulty in trying to identify band crossing visually. The larger graph appears to show a crossing at $\xi = 18$; however, this is simply due to a lack of resolution. The left insert shows that there is a very narrow gap at this frequency value, while the right insert does indeed show a Dirac point at $\xi = 52/3$. If $M$ were to slightly increase from $2.1 \rightarrow 2.2$, then bands 51 and 52 would separate while bands 53 and 54 cross, for $l=2$ and $\{m_1,m_2,m_3\}=\{6,11,10\}$.  Note that while Eq.~\eqref{eq:ratio_3} is sufficent to allow band crossings, it is not necessary. This ratio can only predict values of $\xi_{\text{cross}}$ that are integer multiples of $(2\gamma)^{-1}$.

\begin{figure}[!ht]  
	\centering
	\subfloat[][]{\includegraphics[width=0.8\columnwidth]{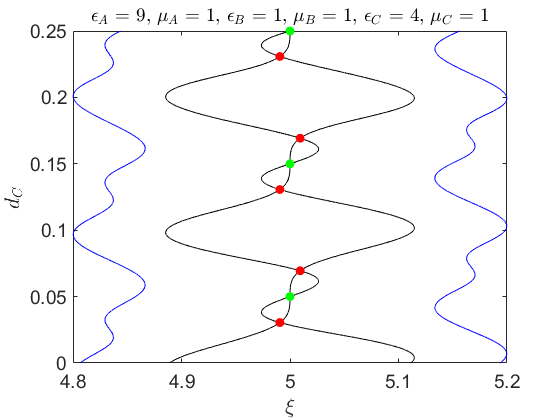}\label{7a}}\\
	\subfloat[][]{\includegraphics[width=0.8\columnwidth]{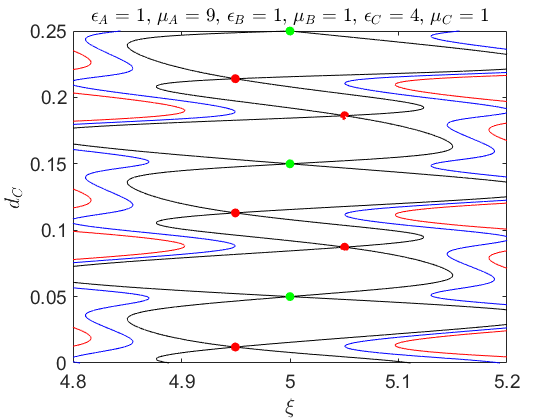}\label{7b}}
	\caption{Locations of the PBG closing points are shown in ($\xi$,$d_C$) parameter space. Blue curve are $t_{11}+t_{22}=0$, black curve are $t_{11}+t_{22}=-2$, and red curves are $t_{11}+t_{22}=2$. Green dots are gap closings described by Eq.~\eqref{eq:ratio_3}; red dots are closings found graphically or numerically.} 
	\label{fig:Zeros}
\end{figure}

The distinction between the crossing points described by  Eq.~\eqref{eq:ratio_3} and those that are not is displayed in Fig.~\protect\ref{fig:Zeros}. The points governed by the ratio are green and occur at integer multiples of $(2\gamma)^{-1}$ (1/3 for these examples). Since the refractive indices in both  Fig~\subref*{7a} and ~\subref*{7b} are the same, these three points remain the same. Note that Fig.~\subref*{6b} is the band structure for Fig.~\subref*{7a} when $d_C=0.05$ ($M=7$). If the band diagram was based on the parameters from Fig. ~\subref*{7b}, it would look slightly different, but the crossing point at $\xi=5$ would remain unchanged. The crossing points denoted in red are not described by Eq.~\eqref{eq:ratio_3}. When $\epsilon_A$ and $\mu_A$ switch values, these points change position, implying that they depend on the impedance.

Another interesting difference between binary and inversion symmetric quaternary PCs is the closing of the first band gap. For a binary crystal, the first PBG will only close when there is impedance matching across the layers of the unit cell (i.e. $z_A=z_B$); however, this matching would cause either $d_A$ or $d_B$ (see Eq.~\eqref{eq:dA} or Eq.~\eqref{eq:dB}) to be negative. Even if layers $A$ and $B$ are forced to have positive widths (by ignoring the constraints of $\Lambda$ and $\Gamma$), every band gap would close, since the PC would exhibit perfect transmission for all frequencies. For the quaternary PC with layer width dictated by Eq.~\eqref{eq:dA} or Eq.~\eqref{eq:dB}, the first gap is allowed to close even if the others remain open. Despite $z_A \neq z_B \neq z_C$, $d_C$ can be tuned so that the lowest gap closes. This closing is not associated with  Eq.~\eqref{eq:ratio_3}, since two of the $m$'s are zero, and thus the associated Dirac point in ($\xi$, $d_C$) space must be found numerically.

\begin{widetext}

\begin{figure}[!ht]  
	\centering
	\subfloat[][]{\includegraphics[width=0.3\columnwidth]{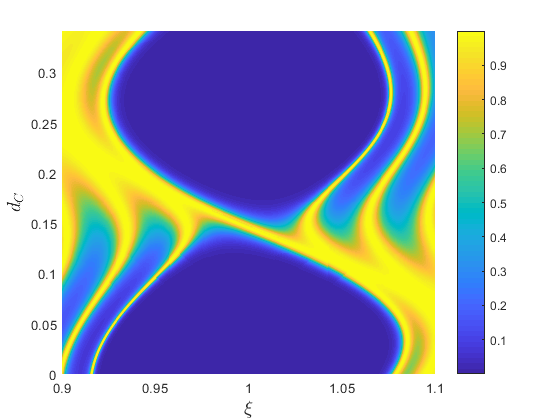}\label{8a}}
	\subfloat[][]{\includegraphics[width=0.3\columnwidth]{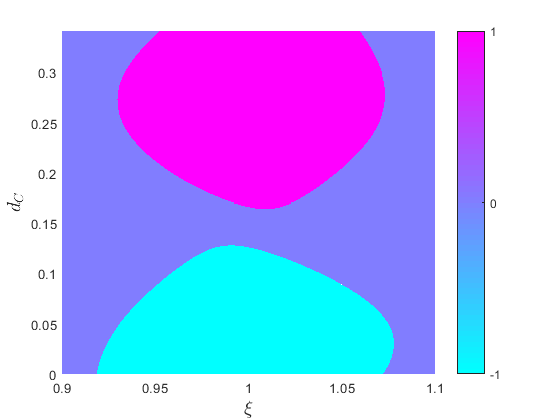}\label{8b}}\\
	\subfloat[][]{\includegraphics[width=0.3\columnwidth]{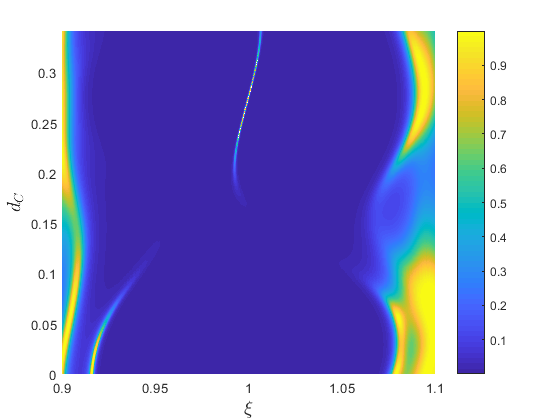}\label{8c}}
	\subfloat[][]{\includegraphics[width=0.3\columnwidth]{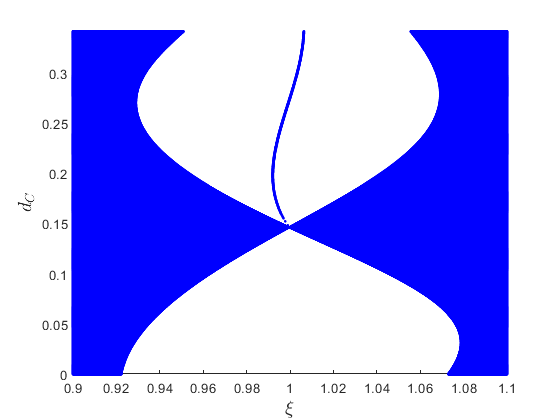}\label{8d}}
	\caption{(a) Transmission spectrum about the $3^{\text{rd}}$ PBG for inversion symmetric quaternary PC with parameters from Fig.~\protect\subref*{2a} as well as $\epsilon_C=3$, $\mu_C=1$. There is one region where the PBG closes then reopens, at $(\xi,d_C) \approx (0.9996, 0.1461)$. Ten unit cells are used. (b) $\text{Sign}\left(\zeta^{(3)}\right)$ is displayed as described by Eq.~\eqref{eq:Zsign}. Colors are kept consistent with  Ref.~\cite{Xiao}.The impedance switches sign after the gaps reopen. (c) Transmission spectrum for combined system of binary and quaternary PCs. For $d_C=0$, the system is described by Fig.~\protect\subref*{1a}.  For $d_C=d^{\text{max}}_C=0.341$, it is described by Fig.~\protect\subref*{1b}. For $0<d_C<d^{\text{max}}_C$, the system is described by Fig.~\protect\subref*{1e}. Five unit cells are used for each PC. Note the topological state after the gap reopens. (d) The state is clearly shown by plotting the implicit equation, $Z_{\text{left}}\left(\xi \right) + Z_{\text{right}}\left(\xi,d_C \right)=0$, where each term is described by Eq.~\eqref{eq:Z_equation}.} 
	\label{fig:InterfaceState1}
\end{figure}

\end{widetext}

\begin{widetext}

\begin{figure}[!ht]  
	\centering
	\subfloat[][]{\includegraphics[width=0.3\columnwidth]{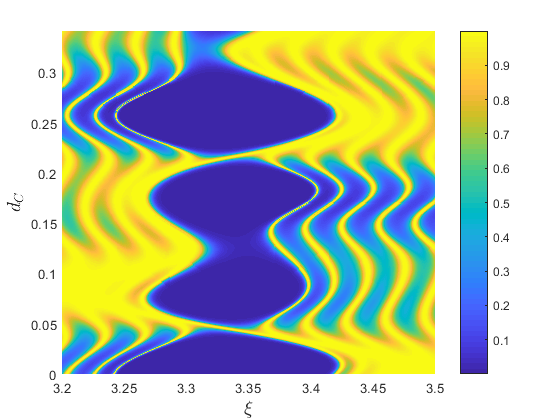}\label{9a}}
	\subfloat[][]{\includegraphics[width=0.3\columnwidth]{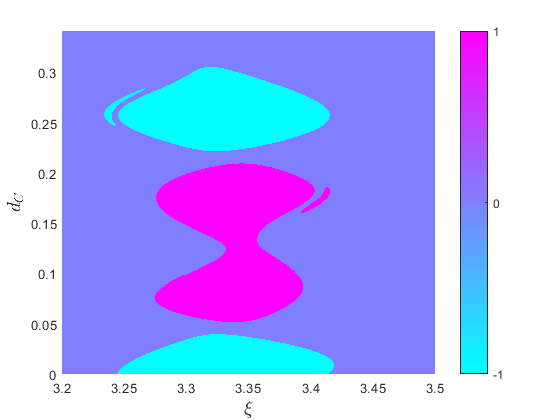}\label{9b}}\\
	\subfloat[][]{\includegraphics[width=0.3\columnwidth]{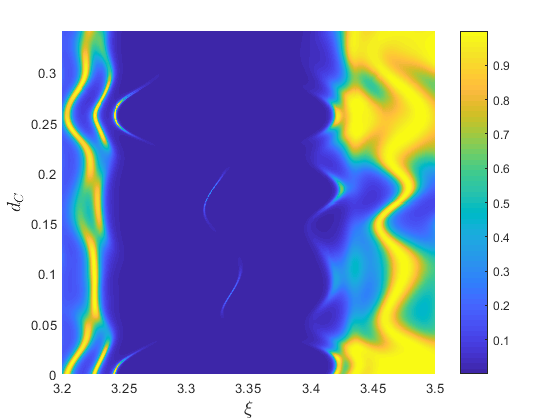}\label{9c}}
	\subfloat[][]{\includegraphics[width=0.3\columnwidth]{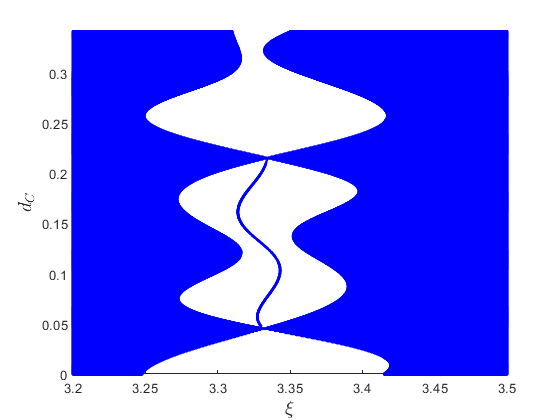}\label{9d}}
	\caption{(a)-(d) Similar to Fig. ~\protect\ref{fig:InterfaceState1} except the plots focus on the $10^{\text{th}}$ PBG. There are two regions where the PBG closes then reopens, at $(\xi,d_C) \approx (3.332, 0.04508)$ and $(\xi,d_C) \approx (3.334, 0.2151)$.} 
	\label{fig:InterfaceState2}
\end{figure}

\end{widetext}

\begin{widetext}

\begin{figure}[!ht]  
	\centering
	\subfloat[][]{\includegraphics[width=0.3\columnwidth]{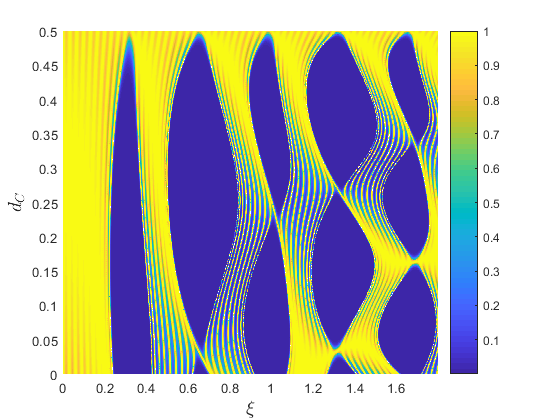}\label{10a}}
	\subfloat[][]{\includegraphics[width=0.3\columnwidth]{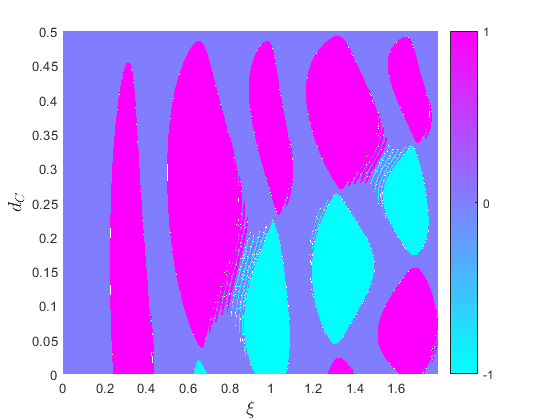}\label{10b}}\\
	\subfloat[][]{\includegraphics[width=0.3\columnwidth]{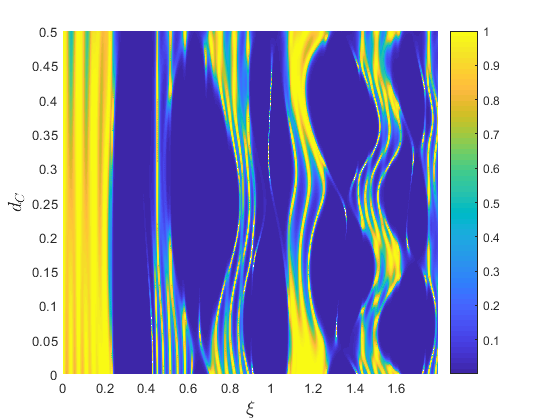}\label{10c}}
	\subfloat[][]{\includegraphics[width=0.3\columnwidth]{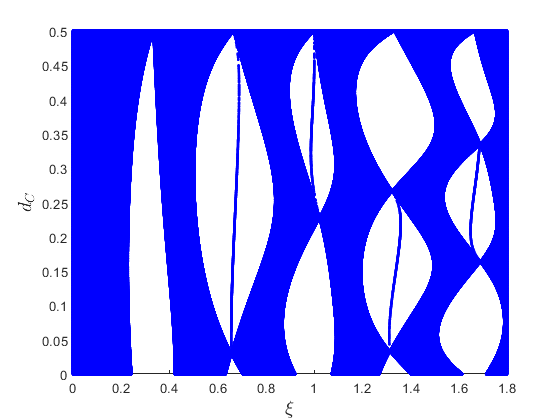}\label{10d}}
	\caption{(a)-(d) Similar to Fig. ~\protect\ref{fig:InterfaceState1} except $\epsilon_C=1$ and $\mu_C=2.25$. PBGs 1-5 are shown} 
	\label{fig:InterfaceState3}
\end{figure}

\end{widetext}

By setting this quaternary PC adjacent to a binary PC, interesting topological behavior is observed. For both PCs, let $\gamma=1.5, ~\epsilon_A=6, ~\mu_A=1,  ~\epsilon_B=1, ~\mu_B=1$ (See  Fig. ~\protect\ref{fig:PC_width}). Three examples of topological state behavior are shown in Fig. ~\protect\ref{fig:InterfaceState1},  Fig. ~\protect\ref{fig:InterfaceState2}, and Fig. ~\protect\ref{fig:InterfaceState3}. For the quaternary PC let the parameters for layer C be  $\epsilon_C=3, ~\mu_C=1$ for  Fig. ~\protect\ref{fig:InterfaceState1} and  Fig. ~\protect\ref{fig:InterfaceState2}, and $\epsilon_C=1, ~\mu_C=2.25$ for  Fig. ~\protect\ref{fig:InterfaceState3}. Let us first consider  Fig. ~\protect\ref{fig:InterfaceState1}. A transmission map about the $3^{\text{rd}}$ PBG of the isolated quaternary PC is displayed in Fig. ~\subref*{8a}, showing two transmission deserts. At $d_C \approx 0.1461$, a Dirac point occurs.  In Fig. ~\subref*{8b}, it is shown that this crossing produces a change in the sign of the surface impedance of the gap, thus producing a change in topological phase of the band structure.  As in Ref.~\cite{Xiao}, cyan is negative impedance and magenta is positive; however, the topology changes due to change in $d_C$ rather than changes in $\epsilon_i$ or $\mu_i$. Ideally, the cyan and magenta parts of the impedance map should meet at a point. The reason why they do not is because the map was created using the transmission map. Everywhere the transmission from Fig. ~\subref*{8a}  was less than some selected percentage (say 0.05), that value would be placed in Fig. ~\subref*{8b} and assigned the correct color according to Eq.~\eqref{eq:Zsign}. Now if the binary PC is placed next to the quaternary crystal, the new transmission is shown in Fig. ~\subref*{8c}. A topological state can be easily seen in the upper half of the map. For the binary PC, $d_C=0$, so the impedance in the $3^{\text{rd}}$ PBG is always negative. As $d_C$ increases in the other crystal, its impedance eventually flips sign. Therefore, in the region of $d_C$ values above the transition, Eq.~\eqref{eq:Zstate} holds and thus a state appears. This is also clearly shown in  Fig. ~\subref*{8d}, in which the imaginary part of Eq.~\eqref{eq:Zstate} is directly plotted. The state can be seen starting from the crossing point. 

In a similar manner to  Fig.~\protect\subref*{8a}, Fig.~\protect\subref*{9a} displays the transmission map for the $10^{\text{th}}$ PBG. The main difference now is that there are two points of band gap closure. Unlike the previous case, the gap width undergoes somewhat oscillatory behavior. It can be seen in  Fig.~\protect\subref*{9b} that the second closing causes the sign of the surface impedance to revert back to the sign it had when $d_C=0$. This means that the topological state produced at the interface between the binary and quaternary PCs will vanish before $d_C=d^{\text{max}}_C$. That state is seen in the transmission map in  Fig.~\protect\subref*{9c}. Lastly, Fig.~\protect\subref*{9d}, clearly shows the interface state starting at the first crossing point and ending at the second. This means that the state only exists for certain intermediate values of $d_C$, for which the superlattice configuration is Fig.~\protect\subref*{1e}. It is not present for the PC configurations corresponding to the extreme values of $d_C$ (i.e. Fig.~\protect\subref*{1a} for $d_C=0$ and Fig.~\protect\subref*{1b} for $d_C=d^{\text{max}}_C$). This is in constrast to  Fig.~\protect\subref*{8d}, where the state persisted for $d^{\text{max}}_C$.

In  Fig.~\protect\subref*{10a} and Fig.~\protect\subref*{10b}, the transmission and impedance maps are shown for a quaternary PC where $n_C=\gamma=1.5$. Note that unlike in the previous two examples where PBGs remained open when $d_C=d^{\text{max}}_C$, all gaps close as $d_C \rightarrow 0.5$, leading to a final configuration shown on the right hand side (RHS) of Fig.~\protect\subref*{1d}. The PC becomes a uniform medium with index, $n_C$. With the inclusion of the binary PC on the left hand side (LHS) of the interface, multiple localized states appear, as seen in Fig.~\protect\subref*{10c}. In total, four are present in the given frequency range, with all them appearing between the first and second gap closing in their respective gap. The states are more clearly defined in  Fig.~\protect\subref*{10d}.

%
\section{Conclusion}
%

The optical properties of inversion symmetric quaternary PCs have been investigated along with the interface states that appear at the boundary of binary and quaternary PCs.  While the quaternary crystal undergoes a smooth transition as the width of the additional layer increases, the band structure undergoes topological phase changes as the photonic band gaps close and reopen, flipping the sign of the surface impedance. Double frequency crossings, which do not exist in binary photonic crystals, are shown to be associated with a Zak phase of 0. The connection between band crossing points and the ratio, M, of trigonometric phase arguments are displayed. As long as this ratio is rational, there will be a band crossing. As the number of non-repeating digits in M increases, the first crossing occurs at increasingly higher frequencies. These band crossings associated with M are only one set of crossings. Other crossings, that have no counterpart in binary photonic crystals, must be found numerically. The evolution of topological states as a function of geometry is explored for cases when the quaternary crystal forms an interface with a binary crystal, and rules for their existence are given. The study of the interaction of multiple interface states in ongoing.

\clearpage

\begin{widetext}

\appendix
\section{Transfer Matrix Method for general quaternary unit cell}

In this appendix, the dispersion relation for a PC with a general 4-layer unit cell is derived. The derivation follows a similar format to the binary unit cell, found in \cite{Yariv}. The direction of propagation is normal to the interfaces in the +$\hat{\textbf{z}}$ direction, with $\textbf{E}=E_x \hat{\textbf{x}}$ and $\textbf{H}=H_y \hat{\textbf{y}}$, where:

\begin{figure}[!ht]  
	\centering
	\subfloat{\includegraphics[width=0.8\columnwidth]{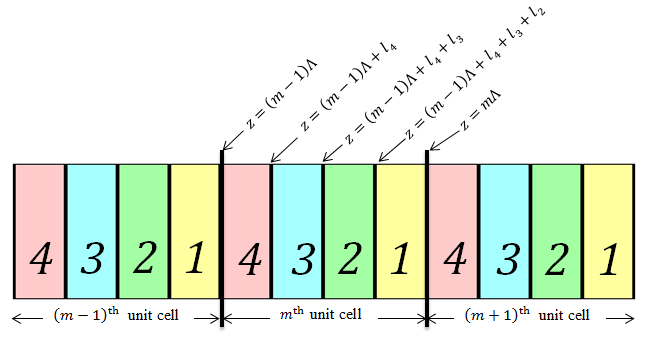}}
	\caption{Diagram of the PC used in the calculations below}  
	\label{fig:4LayerPC}
\end{figure}

\begin{equation}
E_x = a^{(\alpha)}_m e^{-i k_\alpha (z-m\Lambda)} +  b^{(\alpha)}_m e^{i k_\alpha (z-m\Lambda)} \label{eq:Efield}
\end{equation}

\begin{equation}
H_y = \frac{1}{i \omega \mu_\alpha \mu_0}\frac{\partial E_x}{\partial z} \label{eq:Hfield}
\end{equation}
\\*
In Fig.~\ref{fig:4LayerPC}, the unit cell, $m$, is composed of layers, $\alpha = \{1,2,3,4\}$. In Eq.~\eqref{eq:Hfield}, $\omega$ is angular frequency, $\mu_\alpha$ is the relative permeability of layer $\alpha$, and $\mu_0$ is the free space permeability. Matching Eq.~\eqref{eq:Efield} and  Eq.~\eqref{eq:Hfield} across interface $z=(m-1)\Lambda$,

\begin{equation}
\begin{pmatrix}
1 & 1 \\[6pt]
1 & -1\\
\end{pmatrix}
\begin{pmatrix}
a^{(1)}_{m-1} \\[6pt]
b^{(1)}_{m-1}\\
\end{pmatrix}
=
\begin{pmatrix}
e^{i k_4 \Lambda} & e^{-i k_4 \Lambda} \\[6pt]
\frac{z_1}{z_4}e^{i k_4 \Lambda} & -\frac{z_1}{z_4}e^{-i k_4 \Lambda}\\
\end{pmatrix}
\begin{pmatrix}
a^{(4)}_{m} \\[6pt]
b^{(4)}_{m}\\
\end{pmatrix}
\label{eq:layer14}
\end{equation}
\\*
Simplifying  Eq.~\eqref{eq:layer14} yields,

\begin{equation}
\begin{pmatrix}
a^{(1)}_{m-1} \\[6pt]
b^{(1)}_{m-1}\\
\end{pmatrix}
=
\frac{1}{2}
\begin{pmatrix}
\left(1+\frac{z_1}{z_4}\right)e^{i k_4 \Lambda} & \left(1-\frac{z_1}{z_4}\right)e^{-i k_4 \Lambda} \\[10pt]
\left(1-\frac{z_1}{z_4}\right)e^{i k_4 \Lambda} & \left(1+\frac{z_1}{z_4}\right)e^{-i k_4 \Lambda}\\
\end{pmatrix}
\begin{pmatrix}
a^{(4)}_{m} \\[6pt]
b^{(4)}_{m}\\
\end{pmatrix}
\label{eq:Layer14}
\end{equation}
\\*
Matching Eq.~\eqref{eq:Efield} and  Eq.~\eqref{eq:Hfield} across interface $z=(m-1)\Lambda + l_4$,

\begin{equation}
\begin{pmatrix}
e^{i k_4 (\Lambda - l_4)} &   e^{-i k_4 (\Lambda - l_4)} \\[6pt]
e^{i k_4 (\Lambda - l_4)} & - e^{-i k_4 (\Lambda - l_4)}\\
\end{pmatrix}
\begin{pmatrix}
a^{(4)}_{m} \\[6pt]
b^{(4)}_{m}\\
\end{pmatrix}
=
\begin{pmatrix}
e^{i k_3 (\Lambda - l_4)} & e^{-i k_3  (\Lambda - l_4)} \\[6pt]
\frac{z_4}{z_3}e^{i k_3 (\Lambda - l_4)} & -\frac{z_4}{z_3}e^{-i k_3 (\Lambda - l_4)}\\
\end{pmatrix}
\begin{pmatrix}
a^{(3)}_{m} \\[6pt]
b^{(3)}_{m}\\
\end{pmatrix}
\label{eq:layer43}
\end{equation}
\\*
Simplifying  Eq.~\eqref{eq:layer43} yields,

\begin{equation}
\begin{pmatrix}
a^{(4)}_{m} \\[6pt]
b^{(4)}_{m}\\
\end{pmatrix}
=
\frac{1}{2}
\begin{pmatrix}
\left(1+\frac{z_4}{z_3}\right)e^{i (k_3 - k_4) (\Lambda - l_4)} & \left(1-\frac{z_4}{z_3}\right)e^{- i (k_3 + k_4) (\Lambda - l_4)} \\[10pt]
\left(1-\frac{z_4}{z_3}\right)e^{ i (k_3 + k_4) (\Lambda - l_4)} & \left(1+\frac{z_4}{z_3}\right)e^{-i (k_3 - k_4) (\Lambda - l_4)}\\
\end{pmatrix}
\begin{pmatrix}
a^{(3)}_{m} \\[6pt]
b^{(3)}_{m}\\
\end{pmatrix}
\label{eq:Layer43}
\end{equation}
\\*
Matching Eq.~\eqref{eq:Efield} and  Eq.~\eqref{eq:Hfield} across interface $z=(m-1)\Lambda + l_4 + l_3$,

\begin{equation}
\begin{pmatrix}
e^{i k_3 (l_1 + l_2)} &   e^{-i k_3 (l_1 + l_2)} \\[6pt]
e^{i k_3 (l_1 + l_2)} & - e^{-i k_3 (l_1 + l_2)}\\
\end{pmatrix}
\begin{pmatrix}
a^{(3)}_{m} \\[6pt]
b^{(3)}_{m}\\
\end{pmatrix}
=
\begin{pmatrix}
e^{i k_2 (l_1 + l_2)} & e^{- i k_2  (l_1 + l_2)} \\[6pt]
\frac{z_3}{z_2}e^{i k_2 (l_1 + l_2)} & -\frac{z_3}{z_2}e^{- i k_2 (l_1 + l_2)}\\
\end{pmatrix}
\begin{pmatrix}
a^{(2)}_{m} \\[6pt]
b^{(2)}_{m}\\
\end{pmatrix}
\label{eq:layer32}
\end{equation}
\\*
Simplifying  Eq.~\eqref{eq:layer32} yields,

\begin{equation}
\begin{pmatrix}
a^{(3)}_{m} \\[6pt]
b^{(3)}_{m}\\
\end{pmatrix}
=
\frac{1}{2}
\begin{pmatrix}
\left(1+\frac{z_3}{z_2}\right)e^{i (k_2 - k_3) (l_1 + l_2)} & \left(1-\frac{z_3}{z_2}\right)e^{- i (k_2 + k_3) (l_1 + l_2)} \\[10pt]
\left(1-\frac{z_3}{z_2}\right)e^{ i (k_2 + k_3) (l_1 + l_2)} & \left(1+\frac{z_3}{z_2}\right)e^{-i (k_2 - k_3) (l_1 + l_2)}\\
\end{pmatrix}
\begin{pmatrix}
a^{(2)}_{m} \\[6pt]
b^{(2)}_{m}\\
\end{pmatrix}
\label{eq:Layer32}
\end{equation}
\\*
Matching Eq.~\eqref{eq:Efield} and  Eq.~\eqref{eq:Hfield} across interface $z=(m-1)\Lambda + l_4 + l_3 + l_2$,

\begin{equation}
\begin{pmatrix}
e^{i k_2 l_1} &   e^{-i k_2 l_1} \\[6pt]
e^{i k_2 l_1} & - e^{-i k_2 l_1}\\
\end{pmatrix}
\begin{pmatrix}
a^{(2)}_{m} \\[6pt]
b^{(2)}_{m}\\
\end{pmatrix}
=
\begin{pmatrix}
e^{i k_1 l_1} & e^{- i k_1 l_1} \\[6pt]
\frac{z_2}{z_1}e^{i k_1 l_1} & -\frac{z_2}{z_1}e^{- i k_1 l_1}\\
\end{pmatrix}
\begin{pmatrix}
a^{(1)}_{m} \\[6pt]
b^{(1)}_{m}\\
\end{pmatrix}
\label{eq:layer21}
\end{equation}
\\*
Simplifying  Eq.~\eqref{eq:layer21} yields,

\begin{equation}
\begin{pmatrix}
a^{(2)}_{m} \\[6pt]
b^{(2)}_{m}\\
\end{pmatrix}
=
\frac{1}{2}
\begin{pmatrix}
\left(1+\frac{z_2}{z_1}\right)e^{i (k_1 -  k_2) l_1} & \left(1-\frac{z_2}{z_1}\right)e^{- i (k_1 + k_2) l_1} \\[10pt]
\left(1-\frac{z_2}{z_1}\right)e^{ i (k_1 + k_2) l_1} & \left(1+\frac{z_2}{z_1}\right)e^{-i (k_1 -  k_2) l_1}\\
\end{pmatrix}
\begin{pmatrix}
a^{(1)}_{m} \\[6pt]
b^{(1)}_{m}\\
\end{pmatrix}
\label{eq:Layer21}
\end{equation}
\\*
Plugging Eq.~\eqref{eq:Layer21} into the RHS of Eq.~\eqref{eq:Layer32}, Eq.~\eqref{eq:Layer32} into the RHS of Eq.~\eqref{eq:Layer43}, and finally, Eq.~\eqref{eq:Layer43} into the RHS of Eq.~\eqref{eq:Layer14}, 
\begin{equation}
\begin{pmatrix}
a^{(1)}_{m-1} \\[6pt]
b^{(1)}_{m-1}\\
\end{pmatrix}
=
\begin{pmatrix}
t_{11} & t_{12} \\[6pt]
t_{21} & t_{22} \\
\end{pmatrix}
\begin{pmatrix}
a^{(1)}_{m} \\[6pt]
b^{(1)}_{m}\\
\end{pmatrix}
\label{eq:Layer11}
\end{equation}
where the four matrix elements are given by,
\begin{equation}
\begin{split}
t_{11} &= e^{i \phi_1} \big( \cos\phi_2\cos\phi_3\cos\phi_4   
  + i z^{+}_{12}\sin\phi_2\cos\phi_3\cos\phi_4 
 + i z^{+}_{13}\cos\phi_2\sin\phi_3\cos\phi_4 \\
&   -  z^{+}_{23}\sin\phi_2\sin\phi_3\cos\phi_4 
 + i z^{+}_{14}\cos\phi_2\cos\phi_3\sin\phi_4
  -   z^{+}_{24}\sin\phi_2\cos\phi_3\sin\phi_4 \\
&-    z^{+}_{34}\cos\phi_2\sin\phi_3\sin\phi_4 
 - i     z^{+}_{1324}\sin\phi_2\sin\phi_3\sin\phi_4  \big) 
  \label{eq:t11}
\end{split}
\end{equation}

\begin{equation}
\begin{split}
t_{12} &= e^{-i \phi_1} \big(i z^{-}_{12}\sin\phi_2\cos\phi_3\cos\phi_4 
+ i z^{-}_{13}\cos\phi_2\sin\phi_3\cos\phi_4 
+  z^{-}_{23}\sin\phi_2\sin\phi_3\cos\phi_4 \\ 
&+ i z^{-}_{14}\cos\phi_2\cos\phi_3\sin\phi_4 
+   z^{-}_{24}\sin\phi_2\cos\phi_3\sin\phi_4 
+   z^{-}_{34}\cos\phi_2\sin\phi_3\sin\phi_4 \\
&- i   z^{-}_{1324}\sin\phi_2\sin\phi_3\sin\phi_4 \big) 
  \label{eq:t12}
\end{split}
\end{equation}

\begin{equation}
t_{21} = t^{*}_{12}  \label{eq:t21}
\end{equation}

\begin{equation}
t_{22} = t^{*}_{11}   \label{eq:t22}
\end{equation}
where $z_{ij}$ is given by Eq.~\eqref{eq:mismatch1} and $z_{1234}$ is:
\begin{equation}
z^{\pm}_{1324} =\frac{1}{2}\left( \frac{z_1 z_3}{z_2 z_4} \pm \frac{z_2 z_4}{z_1 z_3}\right)
\end{equation}
In Eq.~\eqref{eq:t11} and Eq.~\eqref{eq:t12}, $\phi_i\equiv k_i l_i $. The dispersion relation can be written using the equation $2\cos(\kappa \Lambda) = t_{11} + t_{22}$:
\begin{equation}
\begin{split}
\cos(\kappa \Lambda) &= \cos\phi_1\cos\phi_2\cos\phi_3\cos\phi_4 
   -  z^{+}_{12}\sin\phi_1\sin\phi_2\cos\phi_3\cos\phi_4 \\
& - z^{+}_{13}\sin\phi_1\cos\phi_2\sin\phi_3\cos\phi_4 
   - z^{+}_{14}\sin\phi_1\cos\phi_2\cos\phi_3\sin\phi_4 \\
& - z^{+}_{23}\cos\phi_1\sin\phi_2\sin\phi_3\cos\phi_4 
   - z^{+}_{24}\cos\phi_1\sin\phi_2\cos\phi_3\sin\phi_4 \\
& - z^{+}_{34}\cos\phi_1\cos\phi_2\sin\phi_3\sin\phi_4 
  + z^{+}_{1324}\sin\phi_1\sin\phi_2\sin\phi_3\sin\phi_4   \\
\label{eq:disp1}
\end{split}
\end{equation}
Note that Eqs.~\eqref{eq:t11} - \eqref{eq:disp1} reduce to the expressions for the two and three layer unit cell when $\phi_3=\phi_4=0$ and $\phi_4=0$, respectively. If we let $1 \rightarrow A$, $2 \rightarrow C$, $3 \rightarrow B$, and $4 \rightarrow C$, the dispersion relation reduces to Eq.~\eqref{eq:band_structure}. Alternatively, by using the product to sum identities:
\begin{equation}
\cos\phi_i \cos\phi_j = \frac{1}{2}(\cos(\phi_i - \phi_j) + \cos(\phi_i + \phi_j))
\end{equation}
\begin{equation}
\sin\phi_i \sin\phi_j = \frac{1}{2}(\cos(\phi_i - \phi_j) - \cos(\phi_i + \phi_j))
\end{equation}
Eq.~\eqref{eq:disp1} can be written as:

\begin{equation}
\begin{split}
8\cos(\kappa \Lambda) &=  \left(1 + z^{+}_{12} + z^{+}_{13} + z^{+}_{14} + z^{+}_{23} + z^{+}_{24} + z^{+}_{34} + z^{+}_{1324} \right) \cos(\phi_1 + \phi_2 + \phi_3 + \phi_4) \\
&+ \left(1 -z^{+}_{12} + z^{+}_{13} + z^{+}_{14} - z^{+}_{23} - z^{+}_{24} + z^{+}_{34} - z^{+}_{1324} \right) \cos(\phi_1 - \phi_2 + \phi_3 + \phi_4) \\
&+ \left(1 + z^{+}_{12} - z^{+}_{13} + z^{+}_{14} - z^{+}_{23} + z^{+}_{24} - z^{+}_{34} - z^{+}_{1324} \right) \cos(\phi_1 + \phi_2 - \phi_3 + \phi_4) \\
&+ \left(1 + z^{+}_{12} + z^{+}_{13} - z^{+}_{14} + z^{+}_{23} - z^{+}_{24} - z^{+}_{34} - z^{+}_{1324} \right) \cos(\phi_1 + \phi_2 + \phi_3 - \phi_4) \\
&+ \left(1 -z^{+}_{12} - z^{+}_{13} + z^{+}_{14} + z^{+}_{23} - z^{+}_{24} - z^{+}_{34} + z^{+}_{1324} \right) \cos(\phi_1 - \phi_2 - \phi_3 + \phi_4) \\
&+ \left(1 -z^{+}_{12} + z^{+}_{13} - z^{+}_{14} - z^{+}_{23} + z^{+}_{24} - z^{+}_{34} + z^{+}_{1324} \right) \cos(\phi_1 - \phi_2 + \phi_3 - \phi_4) \\
&+ \left(1 + z^{+}_{12} - z^{+}_{13} - z^{+}_{14} - z^{+}_{23} - z^{+}_{24} + z^{+}_{34} + z^{+}_{1324} \right) \cos(\phi_1 + \phi_2 - \phi_3 - \phi_4) \\   
&+ \left(1 -z^{+}_{12} - z^{+}_{13} - z^{+}_{14} + z^{+}_{23} + z^{+}_{24} + z^{+}_{34} - z^{+}_{1324} \right) \cos(\phi_1 - \phi_2 - \phi_3 - \phi_4) \\
\end{split}
\end{equation}
\end{widetext}


\end{document}